\documentclass[prx,twocolumn,superscriptaddress,floatfix,longbibliography]{revtex4-1}
\usepackage{amsmath,amssymb}

\usepackage{setspace}
\usepackage{graphicx,amsmath,amssymb}
\usepackage{dsfont}
\usepackage{mathtools}

\usepackage{enumerate}
\usepackage[usenames, dvipsnames]{color}
\usepackage{comment}
\usepackage{adjustbox}
\usepackage{graphicx}
\usepackage{subfig}
\usepackage{multirow}
\usepackage{array}
\usepackage{color}
\usepackage{nameref,hyperref}
\usepackage{afterpage}
\usepackage{pifont}
\usepackage{mdframed}
\definecolor{light-gray}{rgb}{0.9,0.9,0.9}

\usepackage{caption}
\usepackage{graphicx} 
\graphicspath{{./Figures/}{./}}
\usepackage{hyperref}
\usepackage[english]{babel}

\usepackage[usenames,dvipsnames]{color}

\usepackage{siunitx}
\usepackage[]{units}
\usepackage{soul}
\usepackage{color, colortbl}
 \definecolor{Gray}{gray}{0.4}

\definecolor{Verde}{rgb}{0.0, 0.5, 0.0}

\begin{document}

\title{Remote teaching data-driven physical modeling through a
  COVID-19 open-ended data challenge.}





\author{Marco {Cosentino Lagomarsino}}
\affiliation{IFOM Foundation, FIRC Institute of Molecular Oncology,
  Milan, Italy}
\affiliation{Dipartimento di Fisica, Universit\` a degli
  Studi di Milano, and I.N.F.N, via Celoria 16 Milano, Italy}
\email{marco.cosentino-lagomarsino@ifom.eu}

\author{Guglielmo Pacifico}
\affiliation{Dipartimento di Fisica, Universit\` a degli
  Studi di Milano, via Celoria 16 Milano, Italy}

\author{Valerio Firmano}
\affiliation{Dipartimento di Fisica, Universit\` a degli
  Studi di Milano, via Celoria 16 Milano, Italy}

\author{Edoardo Bella}
\affiliation{Dipartimento di Fisica, Universit\` a degli
  Studi di Milano, via Celoria 16 Milano, Italy}

\author{Pietro Benzoni}
\affiliation{Dipartimento di Fisica, Universit\` a degli
  Studi di Milano, via Celoria 16 Milano, Italy}

\author{Jacopo Grilli}
\affiliation{The Abdus Salam International Centre for Theoretical
  Physics (ICTP), Trieste, Italy.}

\author{Federico Bassetti}
\affiliation{Politecnico di Milano, Piazza Leonardo da Vinci 32,
  Milano, Italy}

\author{Fabrizio Capuani}
\affiliation{I.N.F.N. Sezione di Roma - Sapienza. Piazzale Aldo Moro 2,
  Roma, Italy}

\author{Pietro Cicuta}
\affiliation{Cavendish Laboratory, Cambridge University, Cambridge,
  United Kingdom}

\author{Marco Gherardi}
\affiliation{Dipartimento di Fisica, Universit\` a degli
  Studi di Milano, and I.N.F.N, via Celoria 16 Milano, Italy}

\begin{abstract}
  Physics can be seen as a conceptual approach to scientific problems, a
  method for discovery, but teaching this aspect of our discipline can be
  a challenge.
  We report on a first-time remote teaching experience for a
  computational physics third-year physics laboratory class taught in
  the first part of the 2020 COVID-19 pandemic (March-May 2020).
  To convey a ``physics of data'' approach to data analysis and
  data-driven physical modeling we used interdisciplinary data
  sources, with an openended ``COVID-19 data challenge'' project as
  the core of the course.
  COVID-19 epidemiological data provided an ideal setting for
  motivating the students to deal with complex problems, where there
  is no unique or preconceived solution. Our results indicate that
  such problems yield qualitatively different improvements compared to
  close-ended projects, as well as point to critical aspects in using
  these problems as a teaching strategy.  By breaking the students'
  expectations of unidirectionality, remote teaching provided
  unexpected opportunities to promote active work and active learning.
\end{abstract}

\maketitle

\section*{Introduction}

Physics is on one hand a corpus of knowledge and a set of technical
and quantitative tools, but on the other hand it is also a conceptual
approach to scientific problems, a way of ``discovering'' that has had
a profound impact on other fields of
science~\cite{Brit2020,nelson2015physical,Hansson2016}.
When we teach a physics class, the corpus may be technical and difficult to
convey, but it is straightforward and usually well defined. For a
particular ``subject matter'', the set of facts and techniques is also
what the students instinctively expect to learn. Conversely, the
flavor for what is a ``physics approach'' is comparatively elusive and
complex to communicate. It is rooted in how we ``do'' physics, how
physicists perform their work in research.
Once data is collected (the act of `measuring' and acquiring data is
the other specialty of physicists) physics digs into this information
through ``physical models'', simple mathematical representations of
empirical observations that have the ambition to capture the essence
of a process. These physics models thus aim to be predictive, i.e.
able to forecast the outcome of independent
experiments~\cite{Brit2020}, ideally also beyond the range used to
inspire and test the model~\footnote{Note that ``model'' is a concept
  used with quite distinct meaning in various fields of science:
  physics, chemistry, biology, engineering. We will use ``model'' as
  shorthand for ``physical model''.}. This approach is extremely
successful within various branches of physics, and today thanks to
detailed datasets from various other fields it also has a great
potential in interdisciplinary and traditionally non-physical science
fields.  Teaching how to use the physics approach productively is
difficult, because it is an unstructured and complex set of skills,
composed of different layers. For many students the acquisition of
these skills is delayed to the first time they are required to produce
original research, such as in their graduation or PhD work, and for
some others these skills are just not acquired.

Certainly what a physicist would call `doing physics' includes many
skills that apply well to all sciences, such as the ability to
formulate hypotheses~\cite{Yanai2019}, and a set of problem-solving
and guess-stimating skills~\cite{swartz2006}, which include being able
to question and ``filter'' information, data, common beliefs, as well
as our own results~\cite{bergstrom2020}. Some of these skills
intersect with what is sometimes called the ``nature of
science''~\cite{Bell2020}, a label that characterizes the research
process as well as the scientific knowledge including socio-cultural
aspects. They also represent what is at the heart of theoretical
science, and one of the most creative and arguably most significant
parts of our work as scientists in general.

During the COVID-19 crisis in Italy, Two of us (MCL and MG) were
teaching for the first time a computational physics laboratory class
for third-year physics students (the other authors of this studies
contributed as students or external supervisors, see below). Our aim
was to convey a ``physics of data'' approach to data analysis and
data-driven physical modeling. We were also committed to using
interdisciplinary data sources rather than conventional physics data
sets, in order to show the students the interdisciplinary potential of
a physics approach to data~\cite{Dolan2015}. The course started in
March 2020, when a national lockdown took effect, so that we had to
rapidly convert the material of the course previously conceived for
traditional frontal live teaching to remote teaching, and we decided
to use ``hot'' data from the ongoing pandemic. What follows is an
account of this teaching experience with two objectives: (1) describe
how the ``hands on'' philosophy described above was formalized and put
in practice and (2) the role of the remote-teaching settings in this
experience.  We found that remote teaching, while being a big
challenge, also offered significant and unexpected opportunities for
this course, leading to effective bi-directional communication with
the students.

The recent literature on computational physics education research is
fairly extensive, due to an original weak integration of computation
into the physics curriculum, as detailed in ref.~\cite{Fuller2006}.
The importance of computation to contemporary physics research is large, as well
its impact on the future employment of physics students beyond physics and
science~\cite{Chonacky2008,Burke2017}. As teachers and researchers
(non-experts of education research), we were interested in a possible
instructive role of open-ended projects (similar to those encountered
professionally) in this context.
The inclusion of such an open-ended part of the course is the main
point of originality of our approach. The account that follows does
not have the ambition of being systematic, but rather we intend this
study to be first and foremost a testimony of our experience, which
can be used to replicate our approach effectively, with an awareness
of its potential and its limitations.
Previous, more systematic, efforts to build courses around expert
practice~\cite{Burke2017} proved to be successful, but were limited to
close-ended projects and exercises. On the other hand, there is
evidence that productive results can stem from learning environments
that problematize topics, put students in charge of addressing problems,
as well as holding them accountable~\cite{Engle2002}.

\begin{mdframed}[backgroundcolor=light-gray]
  \vspace{0.2cm}
\textbf{BOX 1: course essential content.}

\begin{itemize}
\item The \emph{scientific toolbox} of the course (covered in
  approximately 16h of lectures) aims to provide a bare-bone
  scientific set of tools for model-driven data analysis, including
  basic tools from modeling, critical scientific thinking, essential
  probability and statistics and data visualization.

\item The \emph{computational toolbox} of the course (estimated in 16h
  of lectures) includes essential Python tools to treat, analyze and
  plot data (based on the Scipy package, and using NumPy, Pandas, and
  Matplotlib libraries), and typesetting of LaTex documents, useful
  for the reports. We also provided command-line tools to treat,
  analyze and plot data (based essentially on Bash scripting, the awk
  language, and the gnuplot plotting program) and an introduction to
  C++ tools for efficient data analysis and simulations (including
  Standard Template Library data structures).
\end{itemize}
  \vspace{0.1cm}
\end{mdframed}


\section*{Structure and content of the course}

\subsection*{Course layout}

The course fits in the 3rd year physics curriculum at the University
of Milan as a computational physics laboratory class.  It is scheduled
over 66 hours during one semester, which are normally spent ``in the
lab'' with the teaching shared across two instructors. The expected
hours of coursework are approximately 30-40.
%
%
The purpose of the course is to provide (i)~basic notions of
computational tools (C++, shell and scripting languages, python,
latex) and (ii)~skills in  ``physics of data'', i.e. model-guided
data analysis and data visualization, in the form of three short
projects.






The course was structured in a two parts. The first part of the course
(described in BOX 1) is an essential technical and scientific toolbox
for taking off with the projects.  The second part of the course
addresses three one-week long ``data challenge'' individual projects
(DC1, DC2, DC3). In these data challenges, students start from a
dataset and work on it to extract and present conclusions. At the end
of each data challenge, students submit a three-page report written in
LaTeX, to describe the results achieved and to include their own
graphics and figures.  We decided that DC1 would be fully supervised
and guided, with the students working in close contact with the
tutors. DC2 and DC3 would be more autonomous.
To be able to manage the supervision of the students, we restricted
the attendance to a maximum of 25 students (originally the restriction
was also due to the number of available workstations in the classroom).

The data challenges followed the pipeline: ``Get data $\rightarrow$
Clean up data $\rightarrow$ Explore data $\rightarrow$ Model data
$\rightarrow$ Interpret data''. And the students were given the
prescriptions to take into account the limits (and possible biases) of
the data, to prefer models with fewer parameters, and to be strict and
question every interpretation. They were also warned about the
relevant cultural problems that emerge in interdisciplinary
applications of physics there, where it is necessary to acquire some
specific knowledge of the domain, to be humble and trust the experts
of the field.

\subsection*{Conversion to remote teaching.}

Soon before the course started in early March 2020, it became clear
that it would have to be delivered remotely.  We decided to use a Slack
workspace and Zoom meetings. Lectures were mostly asynchronous,
delivered as posts on the Slack space, which students had to go
through on their own. Lectures were posted on Slack using text and
externally linked material (example code, lecture notes, external
resources).  For example, the YouTube channel ``Calling Bullshit in
the Age of Big Data''~\cite{bergstrom2020} was used as a complement
for several aspects of statistics, data visualization, Fermi
estimates, etc. but we also gave the students our own lecture notes
and handouts on all these topics, including exercises and example
codes to plot data, perform fits (see BOX~1 and the Supplementary Appendix,
providing a compendium of the course).

After each set of lectures, we organized a ``Questions and Answers''
session (Q\&A) on Zoom, where we discussed the material with the
students. Next to the material, we also posted guided exercises on the
computational and scientific parts of the course. In each Q\&A
session, besides leaving the meeting open to any questions, each
student was interviewed and requested to update the instructors on the
study work and exercises that he/she had already done, whether she/he
encountered any difficulty, and was asked to state her/his plans for
the coming days. These sessions lasted 2-3 hours and took place once
or twice a week dependiong on the course development.
The interviews took place over a consecutive three-hour slot and each
student had 5-15 minutes where he/she could present the work done on
the asynchronous material and the coding exercises performed in the
past few days. Starting from the these reports, the instructor would
address the questions and common problems encountered in the study,
and any barriers encountered in the coding or in the conceptual
implementation of the exercises.


\begin{figure}[htb]
  \includegraphics[width=0.48\textwidth]{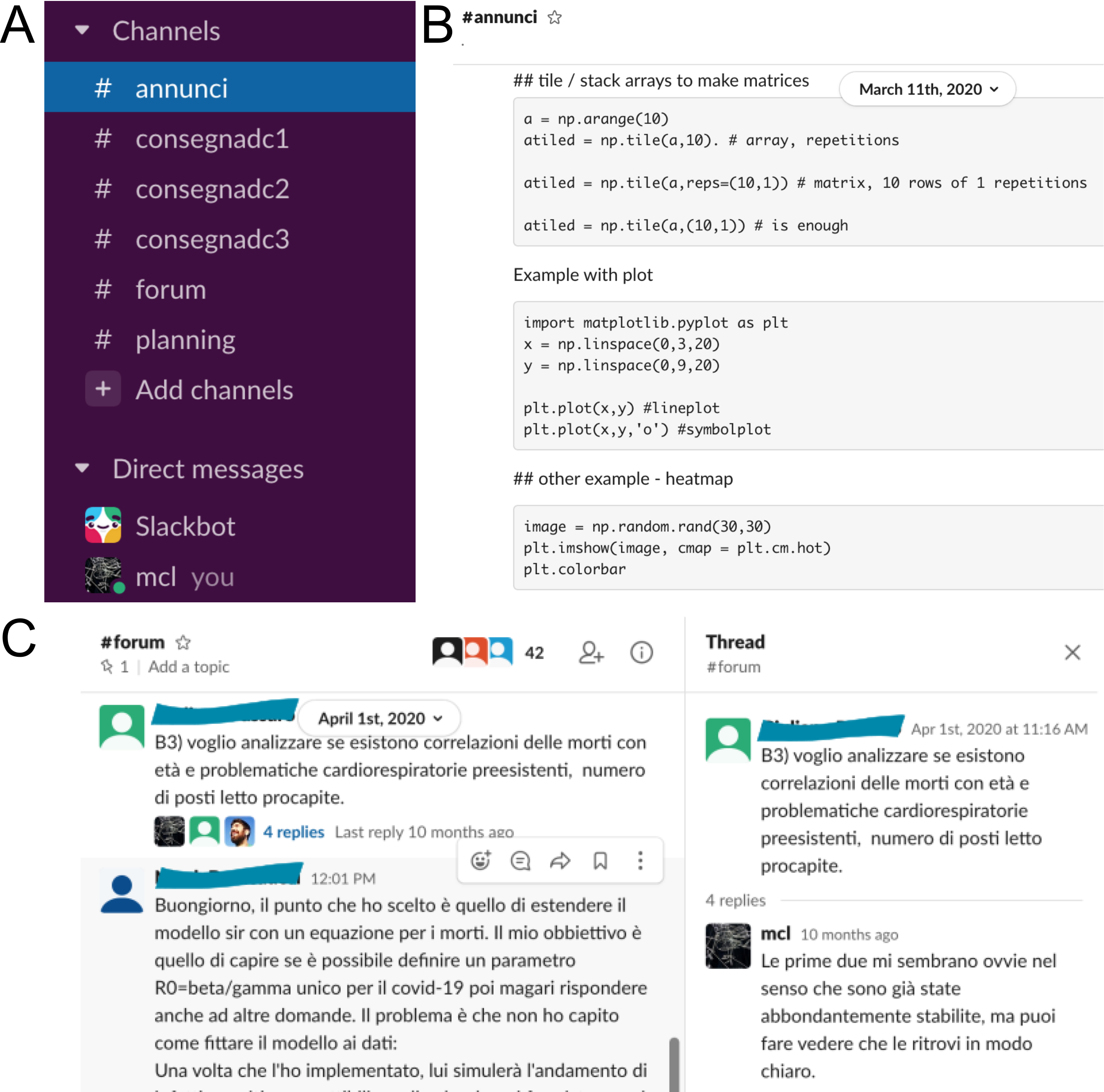}
  \caption{The Slack space facilitates organization of the course,
    posting of lecture material and bi-directional communication.
    A. Subdivision of information into channels for planning of the
    lectures and Q\&A sessions, posting of lecture material, turning in
    data challenge reports. Channels are accessible from a side pane
    of the PC app and as a menu for the mobile app.  B. Example of a
    portion of a posted lecture on Python, with highlighted
    code. Through the interface, a lecturer can include attachments in
    several formats (code, pdf, images, etc.) and links to external
    websites, which the app automatically shows in pretty-print
    preview.  C. Example of the interactive “forum” channels where
    teachers could post external material, and the students could post
    questions publicly. Questions can be addressed in a connected
    “thread” conversation (right-hand panel).  }
  \label{Fig:slack}
\end{figure}

The Slack space was an efficient way to interact bidirectionally with
the students throughout the week (Fig.~\ref{Fig:slack}).
We structured it in several channels for posting teaching material,
planning of the course, receiving material from the students (e.g. the
reports from the data challenges), and we included a forum where
instructors, supervisors and students could post material, tips,
pointers to external materials and data sets, problems and general
questions. The forum turned out to be a very active platform during
the course. One-to-one chats were used and proved efficient for
replying to specific questions. Slack includes an App for cell phones,
which we found useful for real-time communication (and depending on
the teacher’s availability, it can be silenced for prescribed time
periods, whilst avoiding email clutter).

Most students had the right resources to take the course from
home. Five students dropped out when we announced the switch to remote
or during the course. We contacted them through the Slack and email,
but it was difficult to reconstruct the precise reasons (e.g. pressure
from other courses or lack of equipment). We provided instructions
to install all necessary code at home, but each student could also
connect remotely to Linux-based university machines where all the
softwares were available. All codes, software, compilers
supported by the course (essentially, Python, C++, gnuplot, shell
scripting and LaTex for typesetting) were free, although the students
were left free to adopt any tool and judged solely by the end results
of their work.

\subsection*{The COVID-19 data challenge.}

With the pandemic expanding globally, we decided that the first data
challenge would be a COVID-19 data challenge.
Instead, we kept the second data challenge as a more standard
close-ended (interdisciplinary) physics project on multiplicative
diffusion processes, which used data on software
packages~\cite{Gherardi2013,Mandra2014}.  As the course got under way,
we saw that DC1 had (perhaps in hindsight not surprisingly) taken a
lot of time and commitment on both student and teaching staff.
Extensive feedback was being provided to the students on their
projects DC1, so to provide an outcome for this we switched our
original plan and made the third data challenge, DC3, optional, and we
decided that it would consist of an revision and integration the work
carried out in DC1, based on the feedback received from the
instructors. This enabled students to build on the detailed feedback
provided by the instructors on the student reports for DC1. 14 out of
21 students chose to address DC3.
The feedback to students included a number of positive propositions to
expand and revise aspects of each project.

We planned the first data challenge, DC1, to be a supervised project,
so the students were free to ask questions and seek advice on any
level. Since the course started during the COVID-19 crisis (and during
the complete lockdown) in Lombardy and Italy, we thought that putting
their hands on the surge of data that were discussed every day all
over the news and the internet could be both motivating and
challenging.  As mentioned above, in order to supervise the projects
more efficiently during that week, we obtained the help of four
(volunteers) external ``supervisors'' for DC1 (FB, FC, PC, JG,
co-authors of this study).
These scientists provided some ideas for the projects and interacted
directly with the students, individually and through the Slack
forum. The supervisors were based in other cities in Italy and in the
UK, but the Slack interface made it easy for them to interact with the
students through the forum and individually, and to help them carry
out their individual projects.

The external supervisors had different areas of expertise, statistical
and interdisciplinary physics (FC,JG,PC), probability and statistics
(FB), experimental physics (PC) and data analysis (FB,FC,PC,JG).  Each
supervisor rovided his individual perspective to the different
problems and contributed defining possible directions of the Data
Challenge assignment, which was produced as a document with
contribution from  instructors and external supervisors.
Students were left free to find their way to the specific potential
supervisors whose background and vision best suited their goals and
perspective. The resulting partitioning of student between supervisors
was even. Each external supervisor was in touch with 2-4 students, and
the main instructors were in contact with 5-8 students.

Part of the instructions provided for the DC1 were tips on where to
get the data, but the students were encouraged to explore autonomously
different data sources. Part of the technical skills that they
acquired from the first part of the course were on how to ``scrape''
data using command line and Python, and how to manipulate data and
assemble an organized data set (for example, using the Pandas and
Numpy packages in Python). As a basis, the students were encouraged to
get the daily data from:
\begin{itemize}
\item the Italian Civil Protection
  agency.~\footnote{https://github.com/pcm-dpc/COVID-19}
\item The COVID-19 Repository at Johns Hopkins University~\footnote{
    (JHU CSSE COVID-19 Data
    https://github.com/CSSEGISandData/COVID-19}\cite{Dong2020}
  \item  Our World in Data \footnote{https://ourworldindata.org/coronavirus}
  \item the EU ECDC web site
\footnote{https://data.europa.eu/euodp/en/data/dataset/covid-19-coronavirus-data}
 \item  national data from the Italian, French and German
ministries of health.
\item the Italian National Institute of Statistics
  (ISTAT)\footnote{http://dati.istat.it/}
\end{itemize}

The challenge was divided into a first point (A) that was common to
all students and a second one (B) that the students could choose. The
common point was the empirical prediction of the ``infection peak''.
The students were asked to find a suitable and efficient empirical
definition of the peak of the infection. The challenge was to provide
an empirical estimate of the peak using data from different countries
and regions, and propose/validate an empirical method to predict the
peak as accurately as possible. The assignment also provided some tips
on the limitations of the data, some caveats on the requested
analysis, and different possibilities to tackle the question.

Point (B) of the challenge was to identify and address a
well-defined question within a theme chosen from a set of the
following proposed options, all of which concern open scientific
problems.
\begin{itemize}
  \item (B1) Fit / analysis using a standard epidemiological (SIR or SEIR)
model, asking whether unified parameters could be defined for the
spreading of the SARS-CoV-2 virus before the
lockdown\cite{Rdulescu2020}.
\item (B2) Use fits from a SIR/SEIR model to quantify the effect
  (measured as change in parameters) of social distancing, lockdown
  and other non-pharmaceutical interventions~\cite{Giordano2020}.

  \item (B3) Empirical correlative analysis. In
the wake of point (A), define some purely empirical observables from the
data such as delay outbreak-intervention time, delay infections /
death etc. evaluate these quantities in various regions (see above,
staying as local as possible) and correlate them with other possibly
interesting covariates such as population density, public transport /
commuting properties, fine dust pollution, temperature, capacity of
hospitals, etc\cite{Stier2020,Ribeiro2020}.
  \item (B4) ``Bullshit calling''
exercise. Find on the web or in the news a scientific claim  made by
a scientist in a pseudoscientific context (e.g. a plot posted on
facebook or twitter, there are various Italian and international
threads) and then challenge, refute or debunk it using data (and models if
necessary).
\item (B5) Explore possible explanations for the very large variations
  of case/fatality ratios across countries and regions (e.g. 14\% case
  fatality ratio in Italy vs 0.6\% in Germany. The case fatality ratio
  is known to be a very bad estimate of the death rate because of
  delays and other factors~\cite{Wilson2020,Kobayashi2020}. A
  circulating hypothesis was that a confounding factor is the age
  distribution: the Italians are older and therefore die more likely
  of the disease. Another common hypothesis (which quickly turned out
  to be the most reasonable) was that the number of cases was
  underestimated. Another possibility (now ruled out) was that a
  strain of the virus has different death rates.
  \item (B6) Spatial spreading. Identify simple
observables to quantify spatial spreading from available data.  Check
whether containment measures have an effect on this observable. Test
if the opposite has happened: when the measures are announced,
everyone jumps on the train and spreads the virus around the
country~\cite{Gatto2020}.
\end{itemize}

\section{Analysis of the course outcome}

\begin{table}[htb]
\centering
 \begin{tabular}{|c | c|}
 \hline
 \multicolumn{2}{|c|}{DC1/3 project classification}\\
 \hline
 Empirical analysis of epidemiological data	&	20/21     \\
 Epidemic model fitting				&	9/21    \\
 Correlation of epidemiological data with covariates 	&	10/21    \\
 Case-fatality ratio				&	3/21    \\
 Spatial Spreading 				&	4/21    \\
 \hline
 \multicolumn{2}{|c|}{DC1/3 productive findings}\\
 \hline
New dataset / Data integration	  &	8/21    \\
Creative use of mathematical models	 &   9/21    \\
Display of technical rigor	 &	14/21	    \\
Careful controls / statistical analysis	 & 9/21    \\
 \hline
 \multicolumn{2}{|c|}{DC1/3 problems encountered}\\
 \hline
Evident technical flaws		  &	4/21    \\
Misplaced/misleading conclusions   &	5/21    \\
 \hline
 \end{tabular}
 \caption{Classification of DC1/3 projects performed by the students.
   Since the projects were open ended, the students could choose to
   engage different aspets and follow routes. The right column reports
   our classification of the projects by keywords, productive
   findings, and class of problems encountered by the students. The
   right column reports the fraction of projects that fall into each
   (non-exclusive) category.}
 \label{tab:DC1projs}
\end{table}

\subsection*{Results of the COVID-19 data challenges}

From the educational viewpoint, the problem setting was conceived to
encourage the students to put critical thinking into practice and
search for original solutions.
Encouragingly, multiple students came up with approaches to the data
that were original and effective.  Perhaps even more importantly, some
students able to reflect critically on their own work, revising and
correcting their own analyses. For example, one student developed a
technique to predict the infection peak from a logistic fit of the
cumulative curve. In the second report for DC2, she decided to perform
a critical analysis of her own proposition, showing that it did not
work well with data. Other students showed creative behavior on the
level of data assembly and data curation. For example, one student
collected NASA data sets on pollution and correlated these data with
the local COVID-19 case fatality rate. Table \ref{tab:DC1projs}
collects our classification and counts of the kinds of project
strategies chosen by the students, the different kind of productive
findings they were able to achieve, and the main kinds of problems
they encountered.

What follows is a list of some scientifically remarkable findings by
the students, some of which are presented in more detail in Appendix
(Figures \ref{Fig:GP} and \ref{Fig:PVE}).
\begin{itemize}
\item The time delay between the first case in a given Italian
  province and the first case reported in Italy correlated negatively
  with provincial mobility estimated from 2011 census data.
\item There was a factor of 1.5-2.5 (varying from city to city) between
  total deaths in March in cities in the Bergamo area and the sum of
  COVID-19 registered deaths plus the average deaths in the three
  previous years in the same month. Assuming the measured 1.2\%
  mortality, one student could estimate that total cases could be up
  to 50 fold larger than the number of registered cases.
\item Compared to a suitable null model, Italian provinces with a
  first-reported new infected are closer to provinces with already
  ongoing COVID-19 outbreaks compared to a null model where new
  infections travel without spatial contraints.
  \item There is a phenomenological (mildly sublinear) power law relating the
number of reported infected at the peak, and its value on the day a
lockdown was put into effect, valid across regional data from China,
Italy and Spain.
\item The date of the infection peak is roughly independent from the
  date of the lockdown measures, using as reference system for time an
  origin when the new cases are 25.
  \item Bad or biased sampling can be spotted by the time constancy of the
ratio of positive/tested individuals.
\end{itemize}

\begin{table}[htb]
\centering
 \begin{tabular}{|c | c|}
 \hline
 \multicolumn{2}{|c|}{DC1/3 improvements}\\
 \hline
 Addressing questions with modesty & 6/14     \\
 Choosing circumscribed questions  & 4/14     \\
 Supported conclusions / appropriate controls & 10/14   \\
 \hline
 \end{tabular}
 \caption{Evidence of achievement of different goals in the revision
   work from DC1 to DC3. The right column reports the three main
   lessons that we aimed to convey to the students. The
   right column reports the fraction of projects where we found
   evidence that the revisions showed improvements towards achieving
   these goals.}
 \label{tab:DC13lessons}
\end{table}

Beyond the scientific results above, which were only a part of the
goals (see below), we believe that the COVID-19 data challenge
conveyed important specific lessons to the students. First, to address
questions with modesty. Physics provides modeling and data analysis
skills, but no knowledge of epidemiology or other disciplines. First
of all, one must be aware that one is not an expert in trying to
figure out simple things from the data. Second, choosing circumscribed
questions, makes it feasible to reach a goal. Third, an important
result can be positive or negative, but in both cases the conclusions
must be argued carefully, and supported with the appropriate
controls. To support the hypothesis that these lessons were (at least
sparsely) appreciated by the students, we went back to the comparison
of the outcomes of DC3 (the revision) compared to DC1 (the original
project report), and we counted the examples where the revisions
appeared to incorporate these lessons. The results, summarized in
Table \ref{tab:DC13lessons} suggest that (as perhaps one might
expect), the third lesson is the easiest to learn, and the second is
the hardest (it is a problem that most professional scientists
struggle with).

\begin{figure}[htb]
  \includegraphics[width=0.48\textwidth]{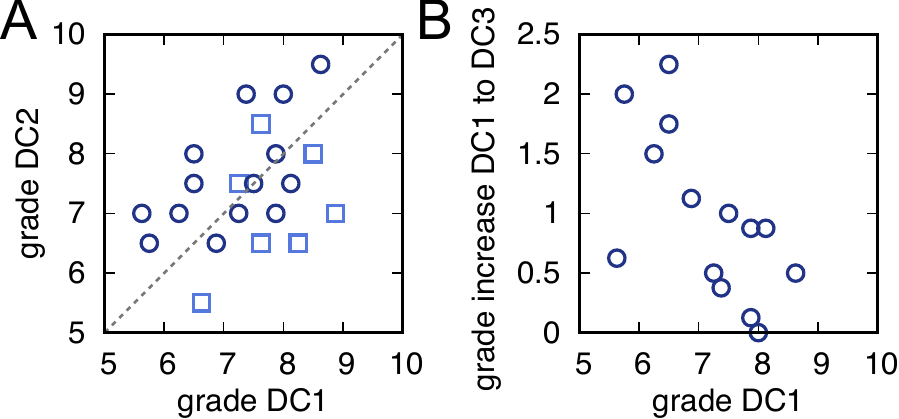}
  \caption{The three-challenge layout allowed comprehensive
    evaluations and offered the students the possibility to increase
    their marks significantly. A. DC1 grades ($x$ axis) and DC2 grades
    ($y$ axis) obtained by all students (circles correspond to those who
    also completed DC3, squares are the others).  B. The net increase
    in grade between DC1 and DC3 ($y$ axis) versus the grade obtained in
    DC1.  Dashed lines are the bisectors $y=x$, as a guide to the
    eye. }
  \label{Fig:grades}
\end{figure}

\subsection*{Grading and evaluation of student improvement}

Ours was a laboratory course, focused on the practical aspects, and
whose cornerstone are the data challenges. It was not trivial to
develop suitable criteria for the evaluation. We wanted the final
grade to reflect the quality of the ``practical'' work done rather
than the acquisition of notions (which can easily be evaluated by an
oral exam). Hence, we decided that the grade should be based on
assessing the reports of the three data challenges.
The reports were graded based on four criteria: (i) Logical structure
and communication; (ii) Data visualization; (iii) Technical aspects of
the analysis; (iv) Scientific aspects of the analysis and support of
the claims.
The benchmarks for a passing grade under these criteria were,
respectively, (i) a clear logical structure divided in sections,
paragraphs, clear result statements and correct reference to
figures, (ii) readability of plots and effective
choice of visual aids, (iii) sound technical choices and controls and
absence of clear technical mistakes (iv) scientifically sound analyses and
adequate support of the conclusions.

\begin{table}[htb]
\centering
 \begin{tabular}{|c | c | c|}
 \hline
 t-tests & DC1 $\rightarrow$ DC3 & DC1 $\rightarrow$ DC2  \\
 \hline\hline
 COM & \textcolor{Verde}{p=0.001} & p=0.8   \\
 VIZ & \textcolor{Verde}{p=0.0003} & \textcolor{Verde}{p=0.03}  \\
 TECH & \textcolor{Verde}{p=0.002} & p=1  \\
 SCI & p=0.2 & p=0.8  \\
 Overall & \textcolor{Verde}{p=0.0004} & p=0.14  \\
 \hline
 \end{tabular}
 \caption{Paired-sample t-tests for the changes in the mean grade of
   students across different data challenges (significant results for
   increased average grades are
   in green). The significant changes from DC1 to DC3 suggest that
   students improved their communication (COM), data visualization
   (VIZ) and technical (TECH) performance in the revision of the COVID-19
   open-ended challenge. The increase in the scientific quality (SCI)
   category was not significant. Conversely, the changes between
   open-ended DC1 and close-ended DC2 are largely not significant,
   likely because of the different nature (and attitude in the
   evaluation) of the project.  }
 \label{tab:ttests}
\end{table}

%
%
Additionally to the grade, we provided extensive feedback for DC1,
from two instructor, in a form similar to a manuscript ``referee
report'' for each student, which included discussion of the weak and
strong points of his/her work and suggestions to correct and/or
improve specific technical, scientific and presentation aspects.  The
final grade was proposed based on the results of all the data
challenges carried out by each student. For those students who did
three data challenges, the final grade was the sum of the three
grades. For those who did two data challenges, we based it on the sum
multiplied by 1.5. We decided that enrolling in the voluntary DC3
could not lower the grade obtained with the first two data challenges,
at worst it would leave it unchanged.  For those who were not
satisfied by the final grade, we made it possible to request an oral
exam. The oral exam could cause both reduction and increase of the
grade based on the students' reports. No student opted for the oral
exam.

Fig.~\ref{Fig:grades} summarizes the grading across the three data
challenges.  Grades received in DC1 and DC2 were correlated only very
mildly (Pearson r=0.39, Fig.~\ref{Fig:grades}A). This suggests that
the two dissimilar challenges allowed us to evaluate complementary
skills and therefore reach a more comprehensive evaluation of each
student. On the other hand, all of the 14 students who addressed DC3
received a grade at least equal to that of DC1, with 5 students
increasing by more than 1 grade (Fig.~\ref{Fig:grades}B).
Moreover, the increase in grade was negatively correlated with the
grade obtained in DC1 (Pearson r=-0.63). This suggests that lower
performing students may have managed to capitalize on the feedback
they received. While lower initial grades leave more room for
improvement, which may explain the correlation, the relative increase
in grade (normalized by the gap to the top grade, which is the maximum
possible increment) still showed weak negative correlation with the
grade obtained in DC1 (Pearson r=-0.27), supporting the hypothesis
(although a larger sample would be needed to establish this point with
confidence).

It is interesting to use the outcomes of the grading to evaluate the
overall impact of the course on the students. The average grades
systematically improved from one data challenge to the next,
suggesting that the students gradually acquired skills to reach the
course goals. Specifically, the average grades (in tenths) for
individual were 7.27 $\pm$ 0.9 for DC1, 7.68 $\pm$ 1.07 for DC2, and
8.12 $\pm$ 0.72 for DC3. To gain more insight on these changes, we
performed t-tests for the increase of the mean grades (Table
\ref{tab:ttests}). We treated the grades as paired samples, regarding
the performance of a student in different data challenges as different
tests of the same criteria in the same subject.
Interestingly, under this test the overall increase of the mean scores
as very significant for the changes between DC1 and DC3 (revisions of
an open-ended challenge), but are largely not significant . Even more
interestingly, students appear to improve their performance from DC1
to DC3 specifically

In support of these results, we also note that the Pearson correlation
between the grades of DC1 and DC2 is 0.1-0.2 (depending on the
evaluator), while the one from DC1 to DC3 is around 0.5 (regardless of
the evaluator).

\subsection{Student evaluation}

At the University of Milan, student feedback  is provided through
anonymous questionnaires based on a set of closed-answers questions,
plus space for open comments.  Generally the questionnaires are filled
by a small fraction of students, but this was not the case for our
course.  We carefully read the answers to the open questions. From
these comments it is clear that the course has aroused ``high
variance'' reactions, both positive and negative.  Multiple students
felt they were thrown into deep waters in a ``sink or swim'' approach,
which disappointed them.  Others were enthusiastic about some aspects
of the course, such as being given an opportunity to develop their
independence, or praising the detailed and constructive feedback that
we provided for their reports, or some specific lectures, such as the
lecture on data visualization (defined by one student as a
``gem''). 5/20 students later on decided to carry out their
Undergraduate thesis projects under the supervision of one of the
teachers.  The large amount of feedback provided by the students is
positive in itself, in the sense that it testifies that the course has
aroused interest and commitment on the part of the students, albeit
with a great variability of starting points in terms of independence
and scientific maturity, which definitely needs to be addressed in the
next editions of this course.

In order to provide more insight into why some students end up
engaging productively and enjoying the realistic scientific setting
science in the course, while others tend to “sink”, we performed two
analyses.

First, we looked at the closed-anwers questionnaires for evidence of
motivation or frustration that could lead to an antagonistic
attitude. We found that 7/21 students thought that the preliminary
knowledge was insufficient, and 12/21 students were not happy about
the material and the mode of the exam (i.e., having to produce a
report for each data challenge). Additionally, 9 students commented
that the charge on the students should be slightly
reduced. Conversely, 19/21 students declared that they were
interested in the topics (13/21 very interested), 14/21 students felt
that the main instructor motivates their interest toward the topic,
but only 3/21 declared that they felt strongly about this.

As a second attempt to gather more evidence, we looked at the
anecdotal evidence from the open questions filled by the students. TO
organize this material, we tried to extract what we found to be key
comments, and relate these comments to a ``sink phenotype'', or a
``swim phenotype''.

SINK PHENOTYPE: \textit{\color{Gray}``I believe that syncronous remote
teaching, for example without frontal lectures where code was
explained in detail, slowed me down a lot in learning the tools
necessary for the DCs'';
``I believe that the frontal lessons are necessary, even if via webcam
or multimedia board'';
``For DC1, the provided material was insufficient to address the
question: to carry out the analysis, it is necessary to know the SIR
model, which was not explained to us except by posting links, and it
is also necessary to know how to integrate a differential equation,
but this has not been explained.'';
``The required computer skills were taken for granted, so much so that
in DC the low familiarity I had with computational tools slowed me down a
lot. I would have preferred to spend more time on the  preparation
part to become familiar with the various tools rather than having to
do it while writing the reports'';
``A negative aspect, which I would recommend to
review, is that often the answers were not exhaustive or clear, in
fact they were often questions themselves.'';
``The course needs more
focus on statistics, models, null models, and hypothesis testing'';
``We students ended up doing a self-taught course; for this, I would
not have enrolled in university'';
``We developed the subjects in an extraordinarily autonomous way. We
had to do all by ourselves'';
``The course has been radically changed from last year, and I only
became aware of it when it started'';
``The partial assessments of DCs were USELESS for the purpose of
having an indication of the final grade''.}

SWIM PHENOTYPE: \textit{\color{Gray} ``The teaching material was
excellent, the grading was clear and above all I appreciated the
mixture of positive and negative remarks in the reports of the DCs'';
``The discovery of the Command Line has opened up a new world for me
and, even if it is as powerful as it is illegible at times, I will
certainly delve into this and other tools that have received only an
introduction in this course'';
``The lesson on DataViz is a gem. I have used very little of the C++,
material, some examples of implementation on simulations would help a
lot to make it feel an integral part of the course'';
``I have learned to write a scientific paper decently, I can work (very
roughly) with a dataset and I have learned many techniques of
data visualization'';
``The course is valid for the acquisition of scientific and computer
skills for the production of a scientific paper'';
``The many computer methods of data analysis covered at the beginning
of the course are interesting, my plotting skills have improved a lot,
even if MatPlotLib becomes a nightmare as soon as you try to make a
custom graph'';
``Really useful course: -Teaches to use programming as a tool to face
and solve concrete problems -Teaches to reflect on the data and on
the results obtained -Teaches the basic principles of scientific
communication'';
``Very useful step for the university journey of a student who follows
this curriculum. Quick exam, useful for the possibility of studying
other exams quickly'';
``The topics covered were interesting and made me discover and deepen
topics that, not being strictly related to physics, I would not have
dealt with in my studies, but I still found them to be important and
useful in a scientific context, for example ``bullshit calling'' and
methods for effective data visualization'';
``However, I realize that it is the main challenge of the course to
provide answers supported by logic and statistics without ever
having used these tools in this way, since the laboratory 1
and 2 courses are bland and the projects are fully guided''.}

In our opinion, these comments fully report on the frustration (of
some students) stemming from feeling abandoned, and also on the
recognition (from some students) of some of the key aspects of the
course.
A parallel consideration is that open-ended challenges open the
opportunity for students and teacher to join sides against common
problems, but this process does not happen automatically for all
students.
We think that an important factor underlying the almost bimodal
reactions is linked to the independence that we asked students to
achieve: one of the training objectives is precisely to make the
student autonomous in learning new tools for analyzing and exploring
data. Before the conversion to remote teaching, the philosophy behind
the course was already to promote active learning through
data-analysis projects.
However, some students, due to lack of information, or simply due to
the fact that the course has radically changed compared to previous
years, expected a more standard course (as witnessed by some of the
comments). This impact should automatically soften over time.

``Inquiry-based learning"~\cite{Brew2003,SpronkenSmith2010,Pedaste2015}
is an approach to teaching that starts from the assumption that a
primary route for a student to learn something is to ask him/herself
questions and then actively look for answers on his/her own (by
contrast classic teaching may give answers to questions that many
would never ask). For this approach, there are various
schools. Typically one tries to engage the students on a problem by
arousing curiosity and stimulating questions.  A ``gradual release of
responsibility'' strategy~\cite{Pearson1983} fits our project, as it
aims at the gradual empowerment of students for their activities, by
showing examples, and gradually ``ramping up'' the scale of difficulty
and independence (which is the most delicate step).

\section*{Discussion}

Unfortunately with the COVID-19 crisis there is likely to be a big
turn away from the experiments in teaching physics in the coming
years, although some initiatives are trying to address practical
teaching suitable for social distancing and remote
learning~\footnote{https://www.smartphysicslab.org/}.  It seems urgent
and useful to set up and systematically improve teaching materials and
experiences such as the one we experimented on here. Such a setting
would lead students to work on real data, analyze curves and
distributions, perform fits, ask questions and look for answers in the
data. It would be a recovery of an important part of what a student
normally learns in laboratory work.

Beyond the contingency of a completely remote learning, the course
offered us an occasion to reflect on general problems related to the
formalization and the implementation of a hands-on approach in a
course aimed at teaching through the supervision of active student
projects.
A full evaluation of the course outcome would require a larger sample,
as well as further efforts towards defining different cohorts of
students and appropriate ``controls'' (for example a quantitative
comparison to an equivalent course with close-ended projects
only). Despite these limitations, our analyses  are in support  the idea
that including open-ended challenges may open the possibility to
address and promote  student skills that are otherwise hard to
access.

Our experience can be replicated in other contexts, not only to
physics students, but also in other quantitative curricula such as
mathematics, engineering or computer science.
A crucial aspect is how many students such a course structure can
support. With two instructors we found that 25 students is a realistic
sustainable upper bound. However, during DC1, the presence of the
external supervisors was essential. These are 10-15 days of the course
where at least 2-3 extra teachers are involved to support the
students' questions and provide feedback. This need should not be
underestimated when planning a similar course, but we believe it would
not be too hard to implement it e.g. through PhD students or postdocs
(also providing a valid supervision experience for early-career
scientists). Not to underestimate, the supervision activity is
probably easier to implement remotely than in a classroom setting. In
such a setting, the extra supervisors can provide feedback through a
chat interface in the time slots and time-frame they want, and the
forum structure leaves a question open to be addressed for any
supervisor (or even other students). It also leaves answered questions
available to any students, who may spot a relevant issue even in cases
when they are not able to crystallize it in a well-defined question.

\section*{Conclusions}

In brief, we judge that the practical implementation of the philosophy
behind the course was successful, but a crucial aspect is to soften as
much as possible the initial barrier for a student to become active on
the ``data challenge'' projects.
Our results support the idea that teaching through open-ended
scientific challenges leads to qualitatively different results than
close-ended projects, and can improve communication and technical
skills, as well as stimulating creativity at different
levels. However, at the same time the same results also point to some
critical aspects of this way of teaching.
Keeping the focus on our main objectives, we intervened in several
ways in the 2021 edition of the course: a) clarifying from the
beginning which skills (soft and hard) we expect the students to
develop, and how they are evaluated in the exam; b) expanding the
description of specific tools, so that students can be immediately
operational; c) providing lectures including ``frontal'' explanations,
combined to the posted study material and the individual Q\&A tutoring
sessions on Zoom, in order to reduce the amount of material to be
processed individually; d) reducing the overall load, passing from
three to two data challenges, and extending the time dedicated to
each, so that students can work under a lower time pressure.
To some extent, the ongoing health crisis helped driving student
interest in the course activities. This fact was still valid in
2021. In the future, to achieve the same level of engagement from
students, the choice of future topics for open-ended data challenges
could be crucial.

From the point of view of Faculty in the physics curriculum, the
constraints enforced by the pandemic led us to innovate on both the
teaching methodology and the subject matter, in ways that we had not
previously explored.  While admittedly far from perfect, we believe
that our experience is something to build on. While the course corpus
and toolbox can be improved, they are nothing special \emph{per
  se}. It is the ``hands on'' part of the course that provided a
valuable access for the students to elements of knowledge, practice
and ``nature of science'' that are typically not accessible in
standard courses, and are often developed through laboratory practical
experiences.  For many, learning ``the physics way'' of addressing
problem was a real hitpoint of the course. Two aspects where
specifically very interesting. First, COVID-19 data was an ideal
setting to learn how to deal with poorly defined ``messy'' problems,
where there is no unique or preconceived
solution~\cite{Dolan2015}. The students had to come to terms with the
exercise of isolating well-defined sub-problems, and to find
compromises of different kinds in order to analyze the data and draw
conclusions, which is what invariably happens outside of a classroom,
in a research or a workplace setting. To this end, the interactive
Q\&A sessions on Zoom where each student could report their findings
and questions were particularly effective. Second, remote teaching
provided unexpected opportunities to stimulate an active role of
students, via oral and written communication. The possibility to keep
a constant communication feedback collectively and individually via
the Slack space was certainly an unexpected benefit for
us. Additionally, and perhaps more importantly, remote teaching breaks
the student expectations of unidirectionality. Looking at a frontal
lecture on line may become extremely boring, and exploring
alternatives is a necessity, but also an opportunity to promote active
work and active learning from the students.



%

\newpage
\onecolumngrid

\appendix*

\section{Examples of students results.}
\label{figurestud}

\renewcommand\thefigure{A\arabic{figure}}
\setcounter{figure}{0}

\begin{figure*}[htb]
  \includegraphics[width=0.4\textwidth]{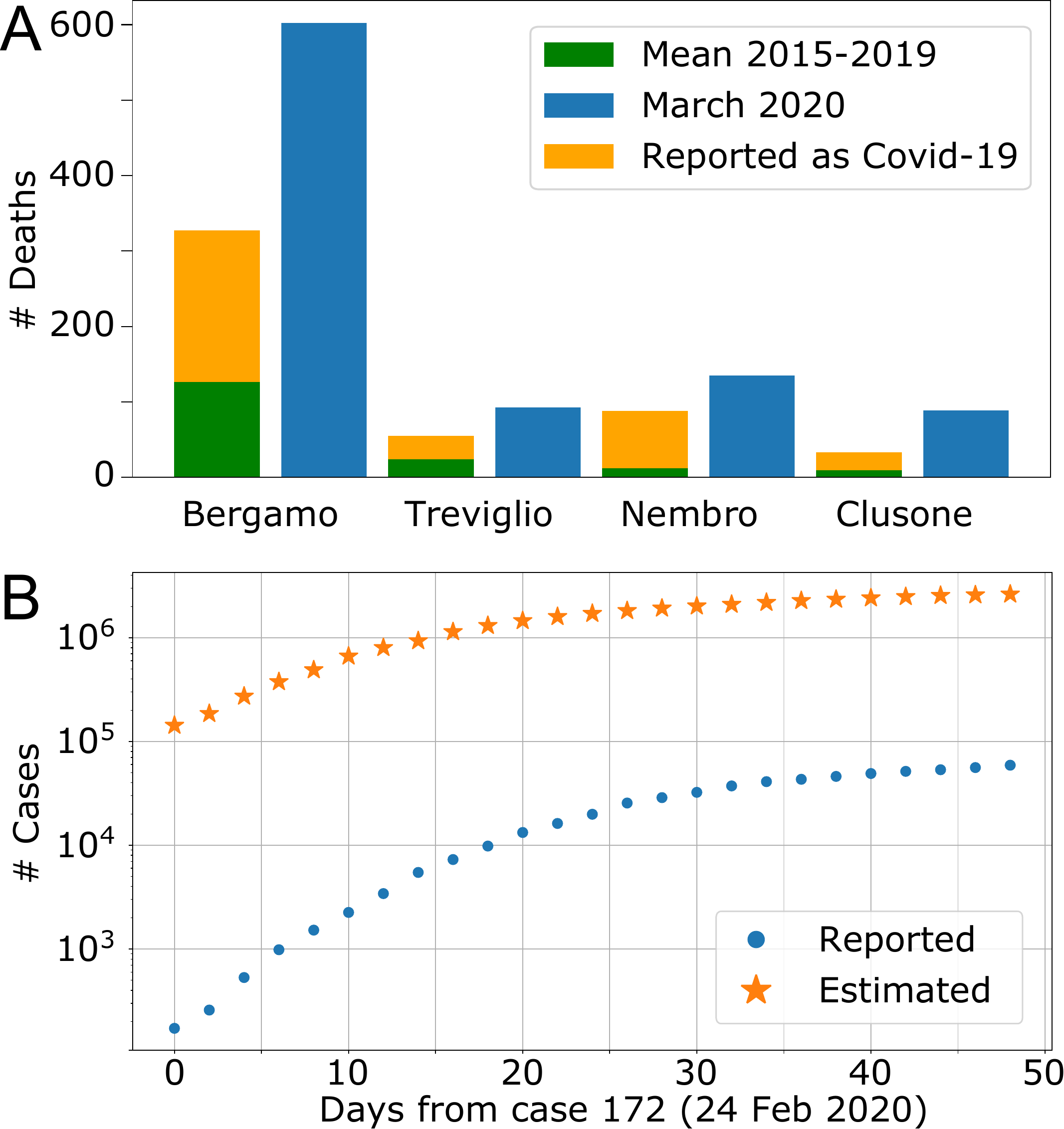}
  \caption{ Examples of students' results on COVID-19 epidemiological
    data combined with mortality data.  Panel A compares the total
    deaths reported in four cities in the Bergamo area in March 2020
    (blue bars) with the sum of the COVID-19 deaths reported in the
    same month (orange bars) and the average of the total reported
    deaths in the same month of the previous five years (green
    bars). In all cases the sum of average mortality and reported
    COVID deaths (green plus orange) was well below the total
    mortality, pointing to widespread unreported cases.  Panel B
    reports an estimate of the actual total number of cases in
    Lombardy (orange stars) from the number of reported cases (blue
    circles) based on the excess mortality. The estimate assumed a
    common overall fatality rate for the virus of 1.2 \% and an
    average delay of 18 days between infection and
    death~\cite{Verity2020}.  Based on these estimates, the actual
    number of cases in that area could have been between one and two
    orders of magnitude higher than estimated from swabs.  2015-2019
    mortality data was downloaded from the Italian National Institute
    of Statistics (ISTAT). Mortality data from the cities in the
    Bergamo area were obtained from the newspaper l'Eco di Bergamo
    (each city had shared these data directly with this
    newspaper~\cite{EcoBerg}). Data were obtained from the Italian
    Civil Protection Repository. }
  \label{Fig:GP}
\end{figure*}

\begin{figure*}[h!tb]
  \includegraphics[width=0.4\textwidth]{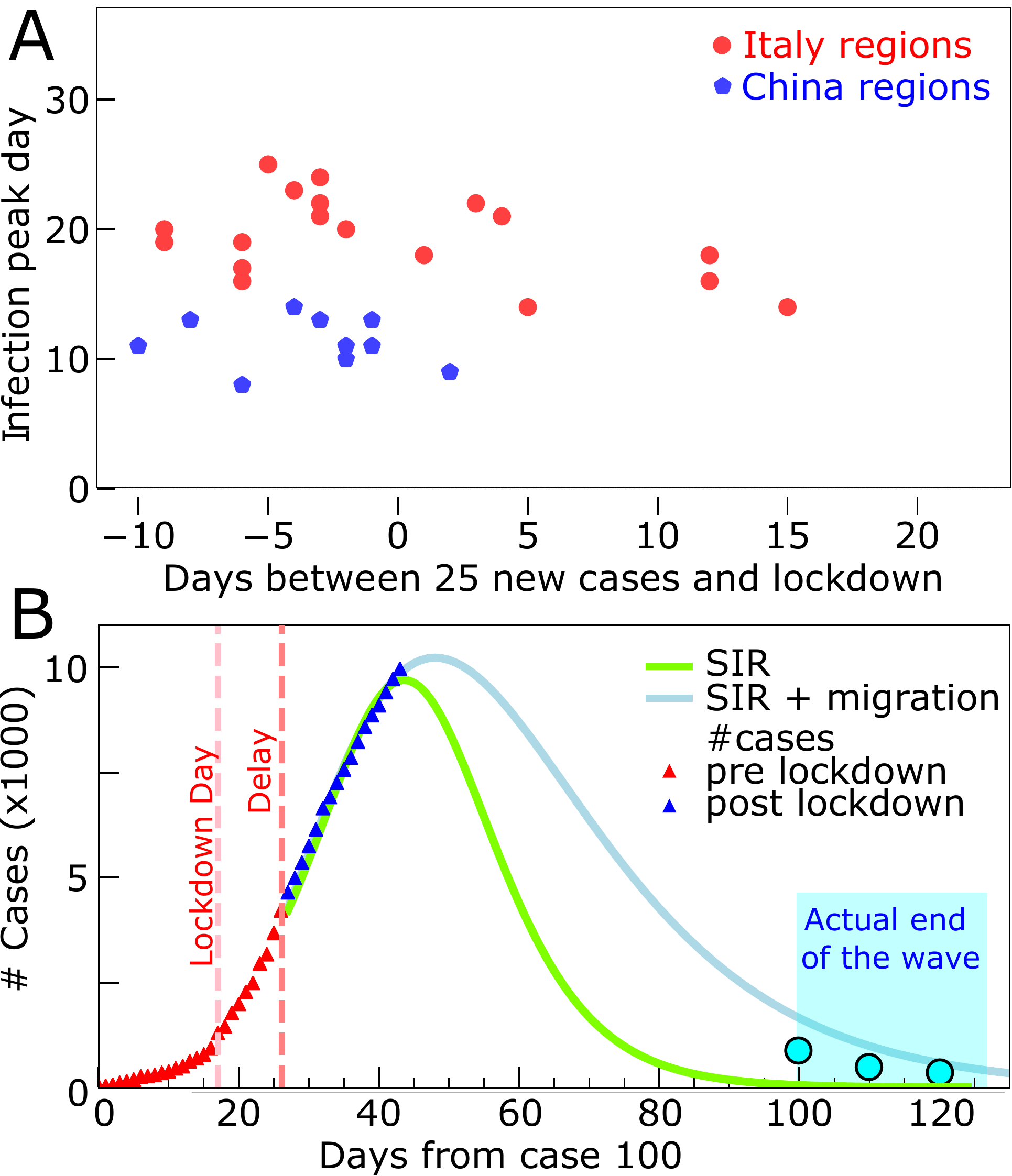}
  \caption{Examples of students' results on COVID-19 epidemiological data.
A. The infection peak, defined empirically as the maximum in the
observed new cases after a lockdown, depends on
the level of the infection at lockdown. The scatter plot reports the
day of the infection peak ($y$ axis, defined as the maximum in the
number of current cases vs the time from lockdown, measured in days
from lockdown) reached by a region compared to the delay between the
implementation of lockdown and the day the number of new cases reached
25 ($x$ axis). Italian regions (red circles) are compared to Chinese
regions (blue pentagons). The plot shows that different regions
took a similar time to reach the infection peak after lockdown, but
the delay was longer in Italy than in China, possibly because of
underreporting or less strict confinement measures.
B. Different infection models lead to different predictions. The plot
shows two fits of SIR disease spreading models performed by two
students, using two different model variants on data from the Veneto
region (triangles). One model (green solid line) was the standard SIR,
while another model (light blue solid line) also included
immigration/emigration. This variant of the SIR model adds parameters
that describe immigration ($\Lambda$) and emigration ($\mu$) and both
variants were used to estimate the effective transmission rate
($\beta$), the hospitalization rate ($\gamma$) and the basic
reproduction number ($R_0= \beta/\gamma$). The model divides the
total population ($N$) into three categories: susceptible ($S$),
infected ($I$), and removed ($R$), with $N = S+ I +R$, and following
the ODEs $dS/dt = \Lambda - \mu S - \beta S I / N$;
$dI/dt = \beta S I / N - (\gamma + \mu) I$;
$dR/dt = \gamma I = \mu R$. The standard SIR model has
$\Lambda = \mu = 0$.
Introducing the model variant lead to a
different prediction of the end of the epidemic wave. The actual data,
which came after the Data Challenge, displayed a behavior in between
the two model predictions. The end of the wave took place about 110
days after lockdown (cyan circles), so that the model with migration
parameters resulted to be more accurate. The model fits kept into
account the fact that the the average number of contacts per unit time
changed after lockdown (blue triangles), compared to before (red
triangles) considering a delay of 9 days for developing the disease.
Data from the the Italian Civil Protection and the JHU CSSE COVID-19
Data Repositories.}
  \label{Fig:PVE}
\end{figure*}

\end{document}